\documentclass{article}

\usepackage[a4paper,margin=25mm]{geometry}
\usepackage{microtype}
\usepackage{graphicx}
\usepackage{amsmath,amssymb,amsfonts}
\usepackage{bm}
\usepackage{physics}
\usepackage{siunitx}
\usepackage{authblk}
\usepackage[numbers]{natbib}
\usepackage[colorlinks=true,linkcolor=blue,citecolor=blue,urlcolor=blue]{hyperref}
\usepackage{caption}
\usepackage{subcaption}
\usepackage{booktabs}
\usepackage{placeins}
\usepackage{xcolor}
\usepackage{indentfirst}
\usepackage{float}

\title{Generation of Deep Ultraviolet Vortices via Amplitude and Phase Spiral Zone Plates\thanks{This document is the results of the research project funded by the Russian Science Foundation (Project No. 23-62-10026)~\cite{RSF}}}
\author{A.S.~Dyatlov\textsuperscript{1,2,\thanks{email:~\href{mailto:aleksandr.dyatlov@metalab.ifmo.ru}{aleksandr.dyatlov@metalab.ifmo.ru}}},
M.A.~Nozdrin\textsuperscript{2},
A.N.~Sergeev\textsuperscript{1},
N.E.~Sheremet\textsuperscript{1},
S.S.~Stafeev\textsuperscript{3},
and D.V.~Karlovets\textsuperscript{1}
}

\affil{\textit{\textsuperscript{1} School of Physics and Engineering, ITMO University, 9 Lomonosova St., Saint-Petersburg, 191002, Russia}}
\affil{\textit{\textsuperscript{2} Joint Institute for Nuclear Research, 6 Joliot-Curie St., Dubna, 141980, Moscow Region, Russia}}
\affil{\textit{\textsuperscript{3} Samara National Research University, Moskovskoe sh., 34, Samara, 43086, Russia}}
\begin{document}

\date{}
\maketitle

\begin{abstract}
    We report the design, fabrication, and demonstration of diffractive optical elements, amplitude and phase spiral zone plates (SZPs), for generating optical vortices in the deep ultraviolet (DUV) range ($260–266\ nm$). 
    Fabricated via photolithography and plasma etching on fused silica, these SZPs efficiently convert Gaussian beams into vortex modes carrying orbital angular momentum. Experiments with a high-power RF photoinjector laser at JINR produced DUV vortices with topological charge $\ell=1$. 
    Comparative analysis shows the phase SZP achieves $\sim 40\%$ conversion efficiency. 
    These results confirm SZPs’ potential for DUV light structuring, enabling applications in electron beam shaping and accelerator technologies.
\end{abstract}

\section{Introduction}\label{sec:Introduction}
    Vortex beams, distinguished by a helical phase front and the carrying of orbital angular momentum (OAM)~\cite{Allen1992}, have emerged as a transformative tool across diverse scientific fields. 
    Their unique properties have been harnessed in telecommunications, quantum information`\cite{Mirhosseini2015}, and optical manipulation~\cite{Perez2023}. 
    Recently, extending these concepts to particle physics has garnered significant interest~\cite{Uchida2010, Verbeeck2010, Blackburn2014}, particularly for generating structured electron beams, which promise new capabilities in materials Fcience and fundamental physics research.

    A primary pathway to producing electron vortices is through photoemission, where the OAM of a laser beam is transferred to the emitted electrons~\cite{Pavlov2024}. 
    For efficient photoemission from robust photocathodes, high-power laser radiation in the deep ultraviolet (DUV) range (4-5 eV) is required. However, generating vortex beams in this regime presents a formidable challenge. Importantly, when generated at higher harmonics, the topological charge of the vortex is multiplied by the harmonic order~\cite{Solomonov2024}.

    Several methods for generating optical vortices have been established, but they exhibit significant limitations when applied to high-power DUV systems. 
    Spatial Light Modulators (SLMs)~\cite{Kumar2024} and Digital Micromirror Devices (DMDs)~\cite{Li2024} are versatile but cannot withstand the high laser fluence typical of photoinjector drive lasers and suffer from performance degradation in the DUV spectrum. 
    Generating a vortex in the fundamental harmonic and subsequently converting it to the fourth harmonic, e.g., DUV, leads to a dramatic reduction in harmonic conversion efficiency due to the complex polarization and intensity profile of the vortex beam~\cite{Sasaki2014}, resulting in an insufficient photoelectron current. 
    Forked diffraction gratings are simple to fabricate but suffer from inherently low diffraction efficiency, as power is split among multiple diffraction orders, typically orders -1, 0, and +1. Spiral Phase Plates (SPPs)~\cite{Oemrawsingh2004} offer high efficiency but their fabrication, requiring continuous or segmented (e.g., 64 steps) spiral topology on DUV-transparent substrates like fused quartz, is technologically complex and costly.

    To address these limitations, we propose and demonstrate the use of Spiral Zone Plates (SZPs)~\cite{Zhang2022, Tsukamoto2022} for the direct generation of DUV vortices. 
    The novelty of this work lies in demonstrating that SZPs, which are significantly easier to fabricate than SPPs, provide a robust and efficient method for structuring high-power DUV laser beams directly at 262~nm. 
    This approach circumvents the power-handling issues of SLMs and the efficiency losses associated with harmonic conversion of existing vortices. This study is part of a larger project to develop a source of relativistic electrons with OAM, a collaborative effort between ITMO University and the Joint Institute for Nuclear Research (JINR).

    In addition to extending SZP-based vortex generation into the DUV spectral range, this work demonstrates the integration of SZPs into a high-power ultraviolet laser system serving as a drive source for a relativistic electron photoinjector at an operational accelerator facility. 
    To our knowledge, this is the first demonstration of vortex beam generation directly within an accelerator laser chain, enabling the transfer of optical orbital angular momentum to relativistic electron beams.

    In this paper, we detail the design and fabrication of both amplitude and phase SZPs. 
    We experimentally validate their performance using the JINR photoinjector drive laser, measure the topological charge of the generated vortices, and compare the conversion efficiencies of the two plate types. 
    Critically, we demonstrate reliable operation under high-fluence, 262~nm conditions typical of RF photoinjector systems, which is a prerequisite for practical OAM transfer to photoemitted electrons and for deployment in accelerator beamlines. 
    Our findings confirm that SZPs are a highly effective technology for future experiments in accelerator physics and quantum electron optics.

\section{Design and Manufacturing}

    SZPs are diffractive optical elements that combine the functionalities of a spiral phase plate and a Fresnel lens~\cite{Sharma2013, Kozlova2019}. 
    This dual nature allows them to simultaneously impart a helical phase front and focus the beam. There are two primary types, which are amplitude and phase SZPs.

    The complex transmission function of an ideal SZP is given by:

    \begin{equation}
        SZP(r, \phi) = \exp \left( i\ell\phi - \frac{i\pi r^2}{\lambda f} \right) 
        \label{eq:mask}
    \end{equation}
    where $(r,\phi)$ are the polar coordinates, $\ell$ is the integer topological charge, $\lambda$ is the operating wavelength, and $f$ is the focal length. The first term in the exponent, $i\ell\phi$, is the characteristic phase of a vortex, which creates the helical wavefront. 
    The second term, $ -i\pi r^2/(\lambda f)$, corresponds to the phase profile of a spherical lens in the paraxial approximation. 
    Thus, the SZP is designed to form a focused vortex beam at a distance $f$ from the plate. 
    The sign of the topological charge, $\pm \ell$, is determined by the handedness of the spiral pattern, that is clockwise or counter-clockwise, which can be controlled during the mask design.

    For fabrication, this continuous function is binarized into a physical mask. 
    The transfer function $T(r,\phi)$ for a binary amplitude SZP is:

    \begin{equation}
        T(r, \phi) = 
            \begin{cases} 
                1, & \sin \left(l\phi - \frac{\pi r^2}{\lambda f} \right) \geq 0 \ \\
                0, & \sin \left(l \phi - \frac{\pi r^2}{\lambda f}\right) < 0 
        \end{cases} 
    \end{equation}
    For a binary phase SZP, which offers higher efficiency, the $0$, or opaque, zones arereplaced with a phase-shifting value of $(-1) \pi$, which corresponds to a $\pi$ phase shift, which is achieved by etching the substrate to a specific depth $d=\lambda/[2(n-1)]$, where n is the refractive index of the substrate material.
    
\subsection{Fabrication Process}

    The diffractive optical elements were fabricated on fused silica substrates, selected for their high damage threshold and transparency in the DUV. 
    A multistep microfabrication process was employed. First, the binary patterns were computationally generated, and a corresponding photomask was produced. 
    A layer of photoresist was then applied to the silica substrate, and the pattern was transferred to the photoresist using standard photolithography techniques to achieve the required resolution.

\begin{figure} [!htbp]
        \centering 
        \includegraphics[width=0.4\linewidth]{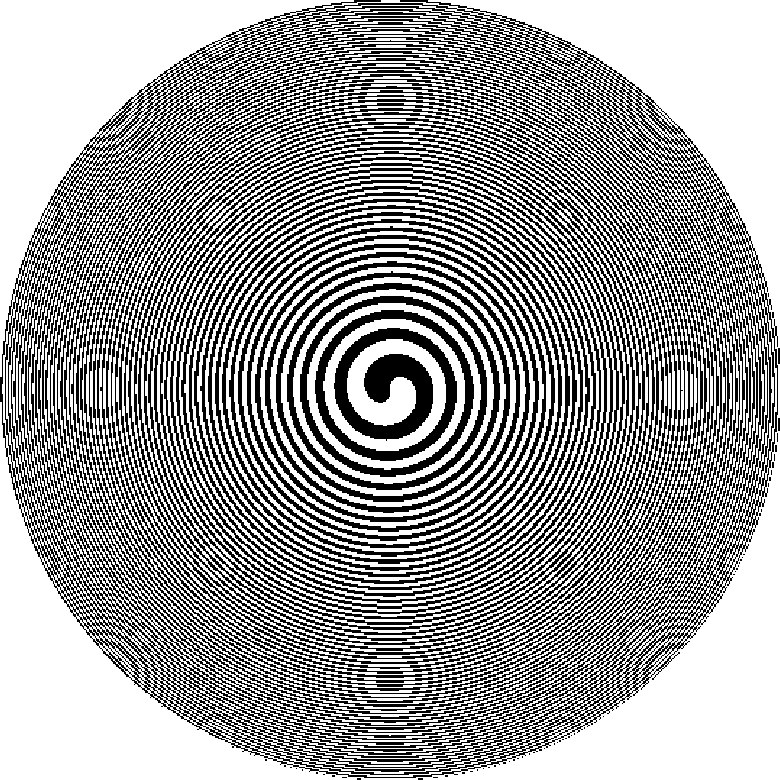} 
        \includegraphics[width=0.4\linewidth]{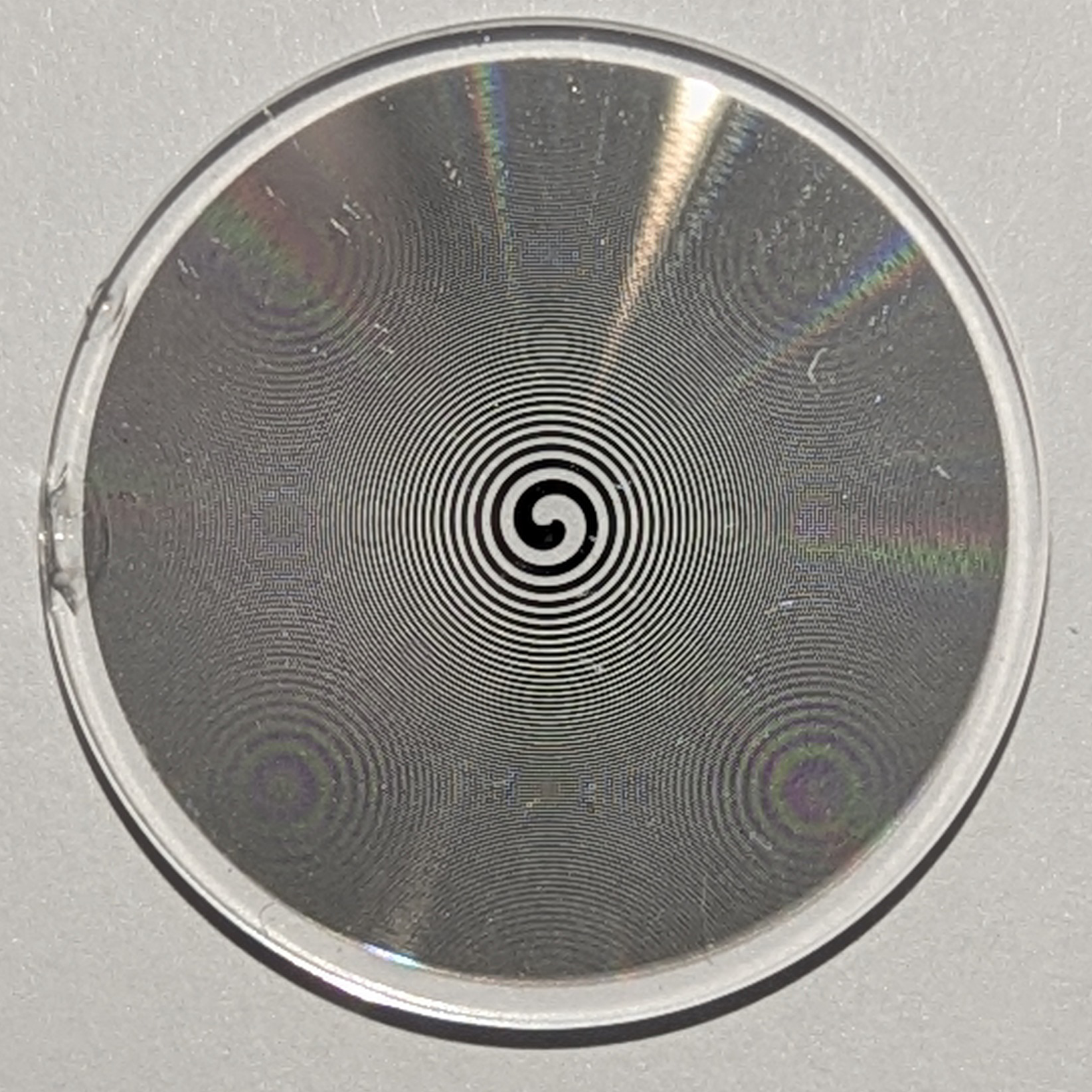} 
        \caption{Fabricated amplitude spiral zone plate. 
        Left: amplitude mask; right: diffractive optical element} 
        \label{fig:amp_szp} 
        \end{figure}

\begin{figure} [!htbp]
        \centering 
        \includegraphics[width=0.4\linewidth]{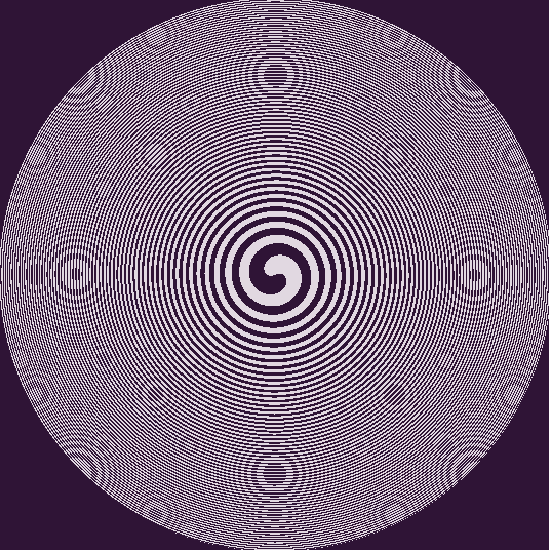} \includegraphics[width=0.4\linewidth]{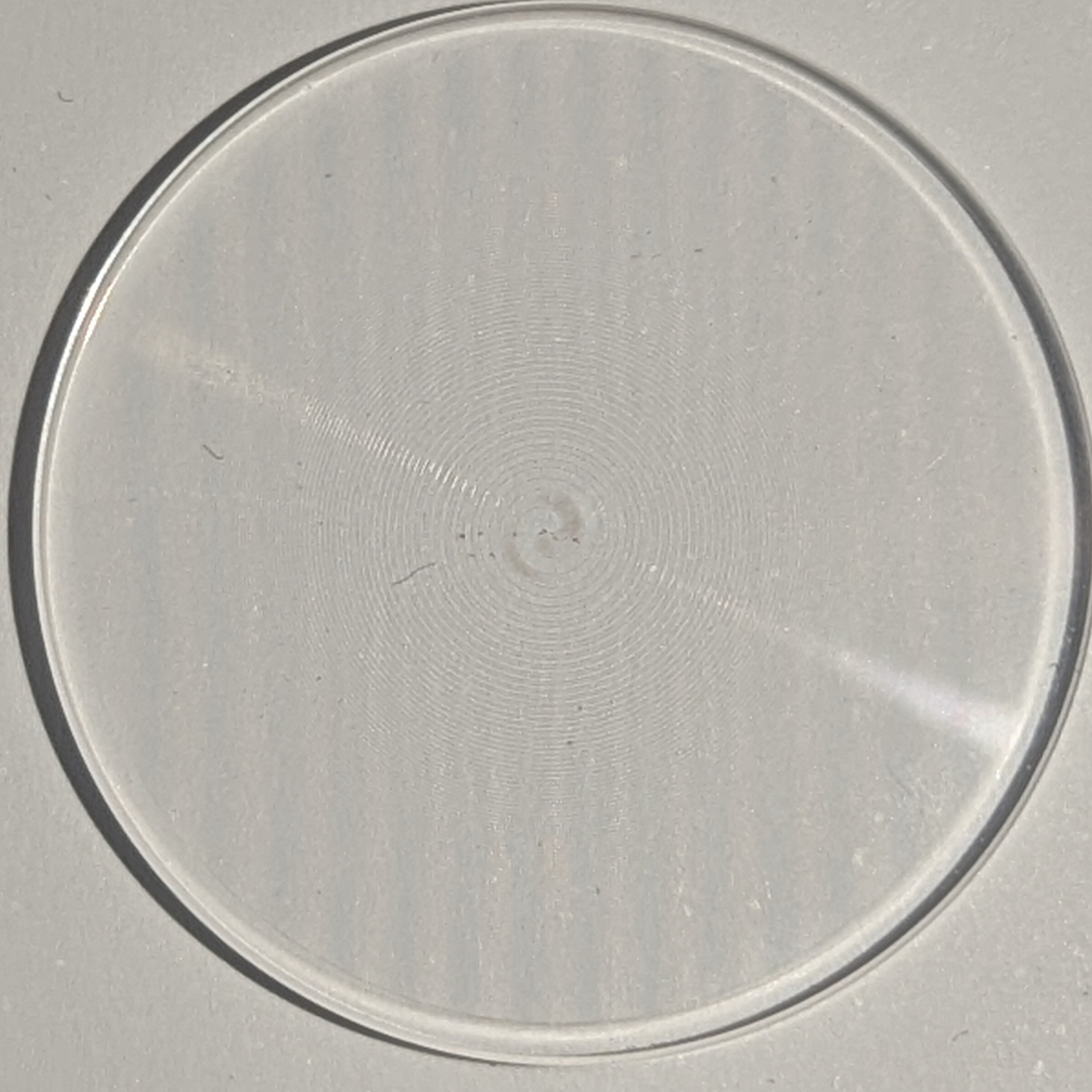} 
        \caption{Fabricated phase spiral zone plate. 
        Left: phase mask; right: diffractive optical element} 
        \label{fig:phase_szp} 
        \end{figure}

    The subsequent steps differed for the two types of plates. 
    For the amplitude plate, a thin, UV-opaque layer was deposited onto the substrate. 
    Following the lithographic patterning, this layer was selectively removed through an etching or lift-off process, leaving the final structure of alternating transparent and opaque spiral zones. 
    For the phase plate, after the pattern was defined in the photoresist, the design was transferred directly into the fused silica substrate using plasma etching. 
    The depth of the etch was precisely monitored and controlled to induce the required $\pi$ phase shift for the operational wavelength, thus creating the phase-modulating relief. In the periphery of the fabricated structures (Figs.~\ref{fig:amp_szp} and~\ref{fig:phase_szp}), weak artifacts are visible. 
    These are attributed to imperfections during the photolithographic pattern transfer stage, rather than the etching process itself.

    The amplitude SZP had a zone period down to $0.6~\mu$m, a working aperture of $5~\text{mm}$, and a focal length of $f = 2500~\text{mm}$. 
    The phase SZP had the same minimal zone period but a smaller aperture of $3~\text{mm}$ and a focal length of $f = 250~\text{mm}$. 
    For the phase element, the required etch depth corresponding to a $\pi$ phase shift at $\lambda = 262~\text{nm}$ was calculated as $d = \lambda / [2 (n - 1)] \approx 267~\text{nm}$, assuming the refractive index of fused silica $n = 1.49$ at this wavelength.

    It is worth noting that the essential distinction between a phase SZP and a spiral phase plate (SPP) lies in the required surface profile. 
    Fabrication of an SPP necessitates producing a continuous helical relief with a smooth spiral surface~\cite{Oemrawsingh2004}, whereas a phase SZP requires only etching shallow grooves of controlled depth into the substrate. 
    This substantially simplifies the fabrication process of SZPs compared to SPPs.

    \section{Experiments and Results}

\subsection{Experimental Setup}

    The experiment was performed using the drive laser of the RF photoinjector at JINR~\cite{Gacheva2014}. 
    The laser system produces pulses at $262\ nm $ by frequency quadrupling a fundamental Yb-doped fiber laser ($\lambda =1047\ nm$). 
    This is achieved using a Potassium Titanyl Phosphate (KTP, $K Ti O P O_4$) crystal for second-harmonic generation and a Beta-Barium Borate (BBO, $\beta-Ba B_2 O_4$) crystal for fourth-harmonic generation.

\begin{figure*} [!ht]
        \centering
        \includegraphics[width=0.9\linewidth]{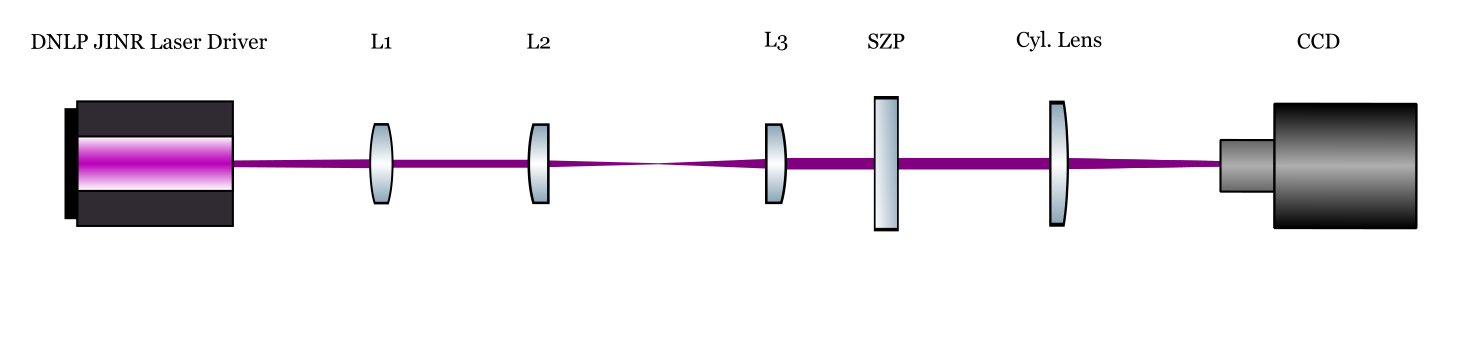}
        \caption{Experimental scheme for vortex beam generation and analysis in the DUV range. 
        The setup includes a collimating lens ($L_1$), a beam expander telescope ($L_2 - L_3$) to fully illuminate the Spiral Zone Plates (SZP) and a mode converter (Cyl. Lens) for analysis.}
        \label{fig:exp_scheme}
    \end{figure*}

    The experimental scheme for vortex generation and analysis is shown in Fig.~\ref{fig:exp_scheme}. 
    The output DUV beam from the laser possessed significant divergence and astigmatism. 
    It was first collimated by lens $L_1$. 
    A telescope system, composed of lenses $L_2$ and $L_3$, was then used to expand the beam. 
    This expansion was necessary to ensure the laser beam diameter was larger than the working aperture of the SZPs, thereby illuminating the entire diffractive pattern for optimal mode conversion. The amplitude and phase SZPs were then sequentially placed in the expanded beam path.

    To analyze the generated beam and measure its topological charge, we used a simplified mode converter consisting of a single cylindrical lens~\cite{Abramochkin1991, Beijersbergen1992, Kotlyar2017}. 
    This well-established technique introduces astigmatism, which breaks the beam's cylindrical symmetry and causes its helical phase front to interfere with itself. 
    This interference pattern manifests as a number of bright intensity lobes (stripes), $N$, which is directly related to the topological charge magnitude, $\vert \ell \vert$, by the simple rule $\vert \ell \vert = N - 1$. 
    A modified CCD camera (SDU-285R), with its protective glass removed to increase DUV sensitivity, was placed in the focal plane of the cylindrical lens to capture the spatial profiles.

\subsection{Results}

\begin{figure} [!ht]
        \centering
        \includegraphics[width=0.49\linewidth]{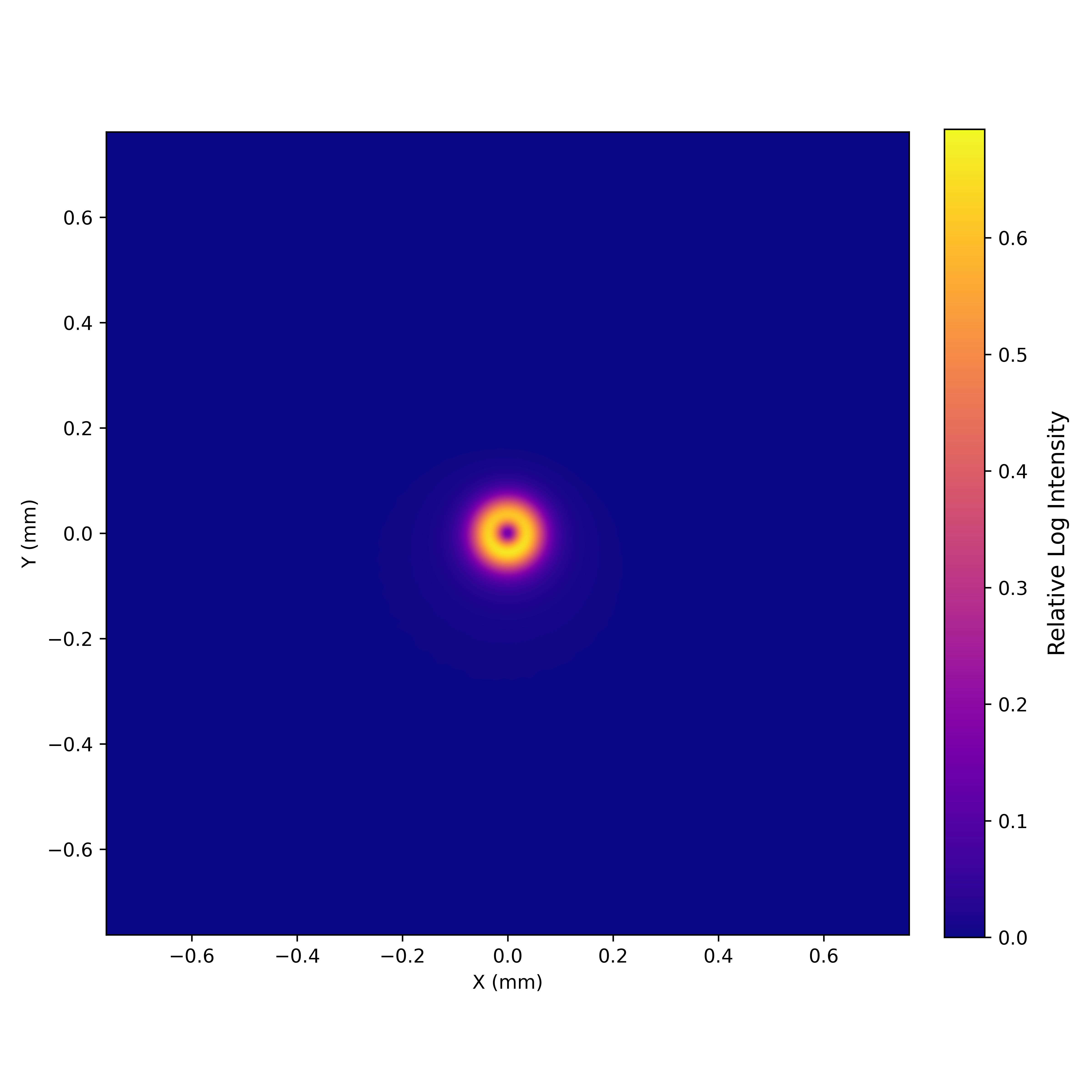}
        \includegraphics[width=0.49\linewidth]{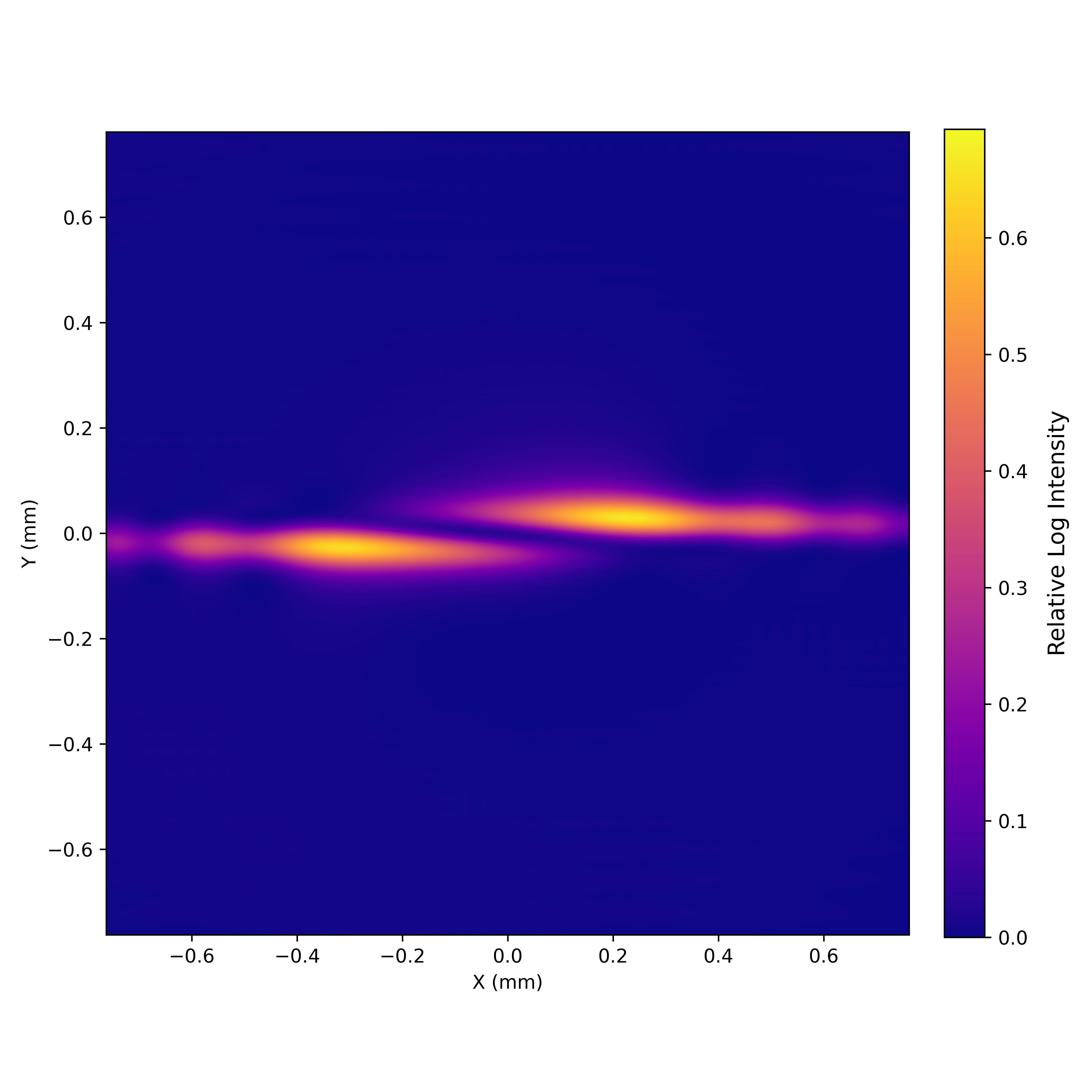}
        \caption{Simulation results of the diffraction experiment at an amplitude spiral zone plate. 
        Left: optical vortices; 
        Right: converted mode obtained after the vortex beam passes through the cylindrical lens.}
        \label{fig:sim_amp}
    \end{figure}

\begin{figure} [!ht]
        \centering
        \includegraphics[width=0.49\linewidth]{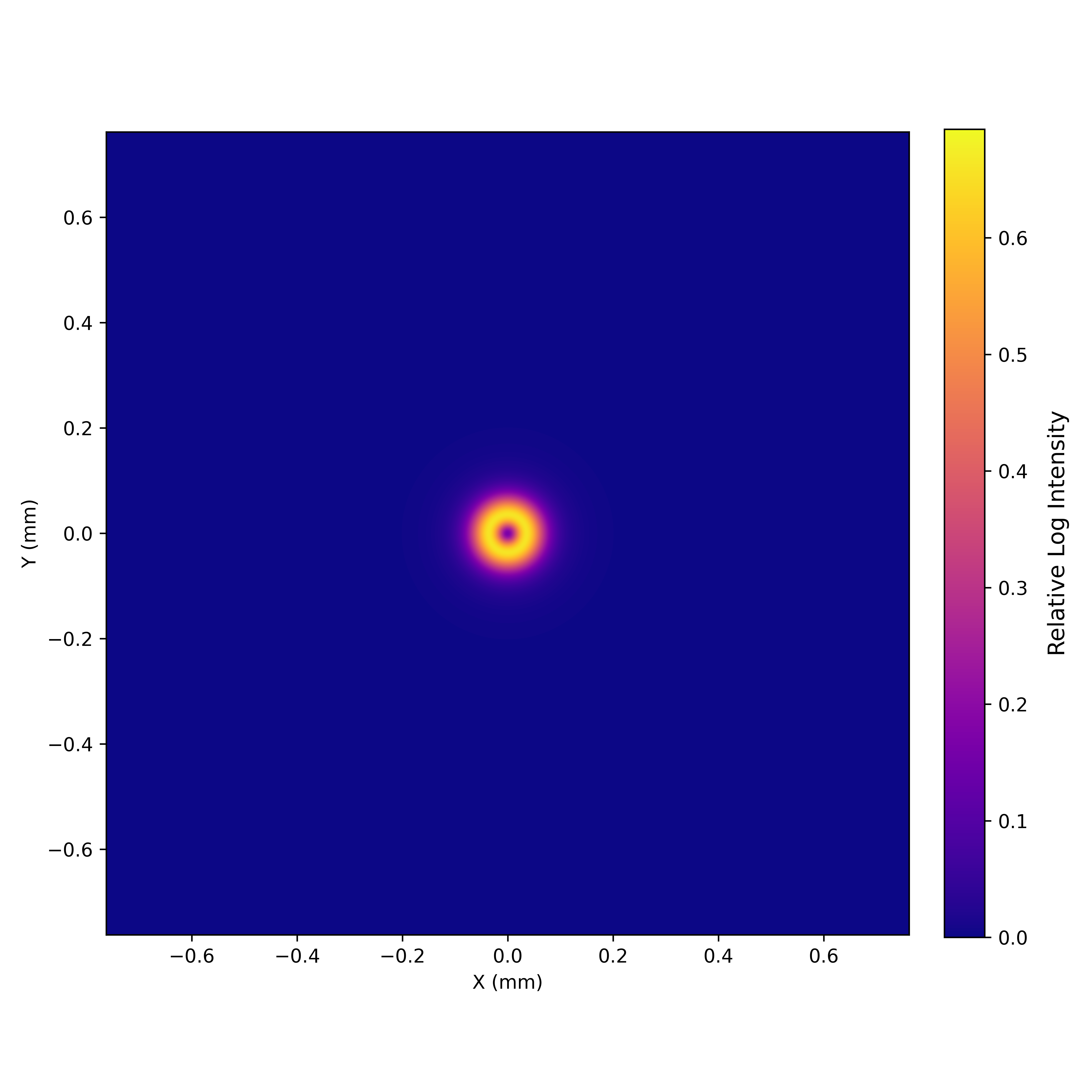}
        \includegraphics[width=0.49\linewidth]{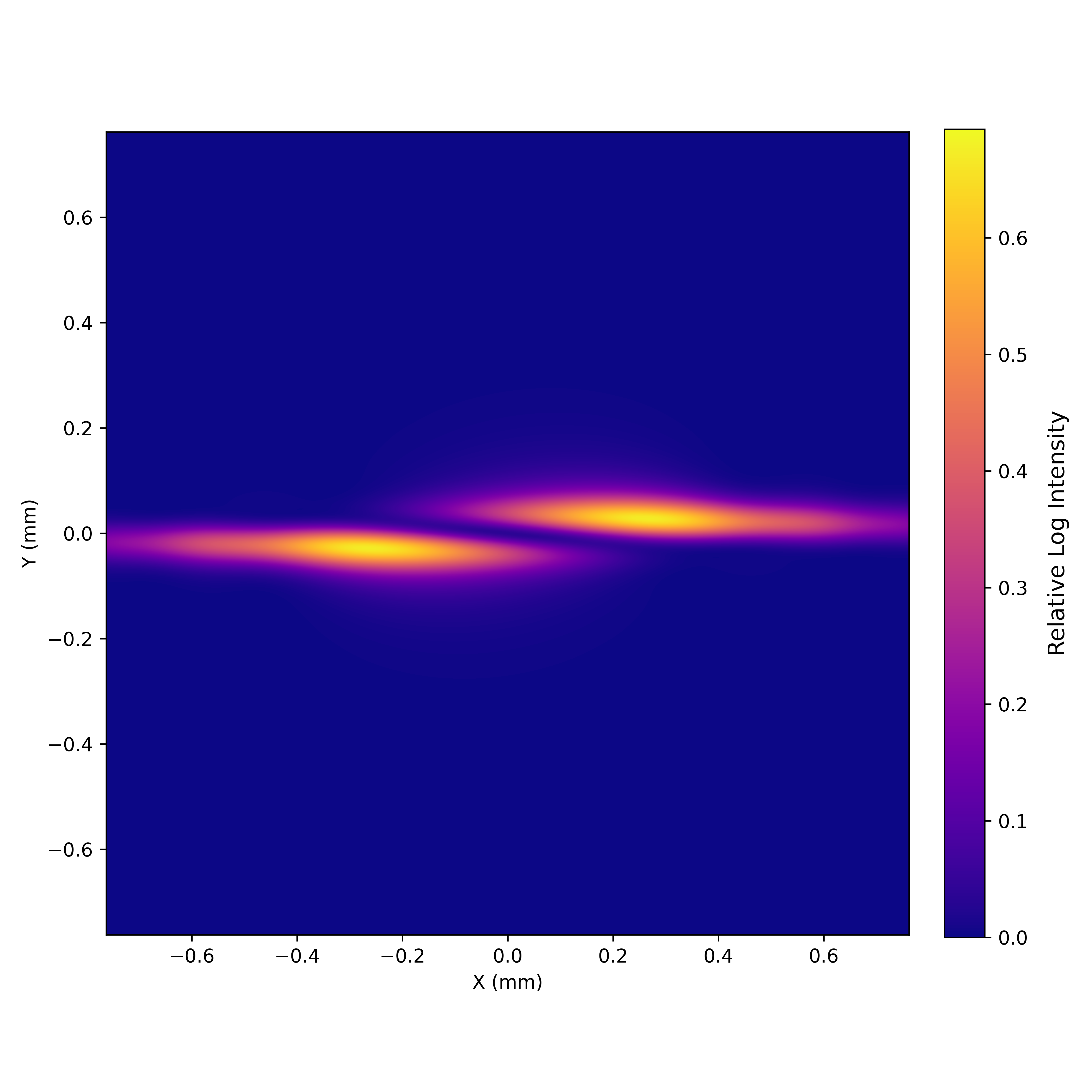}
        \caption{Simulation results of the diffraction experiment at a phase spiral zone plate. 
        Left: optical vortices; 
        Right: converted mode obtained after the vortex beam passes through the cylindrical lens.}
        \label{fig:sim_phase}
    \end{figure}

\begin{figure} [!ht]
        \centering
        \includegraphics[width=0.49\linewidth]{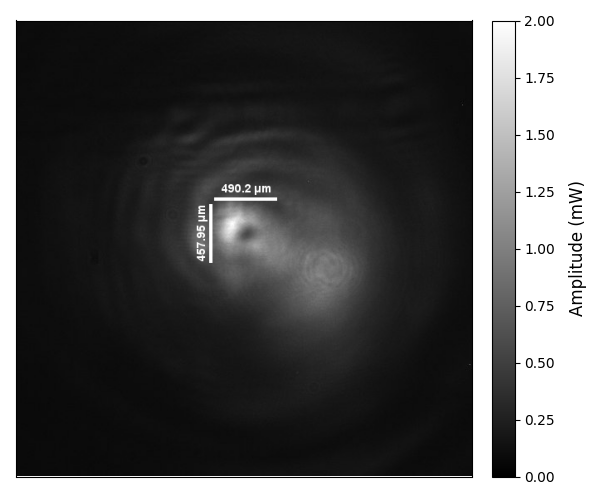}
        \includegraphics[width=0.49\linewidth]{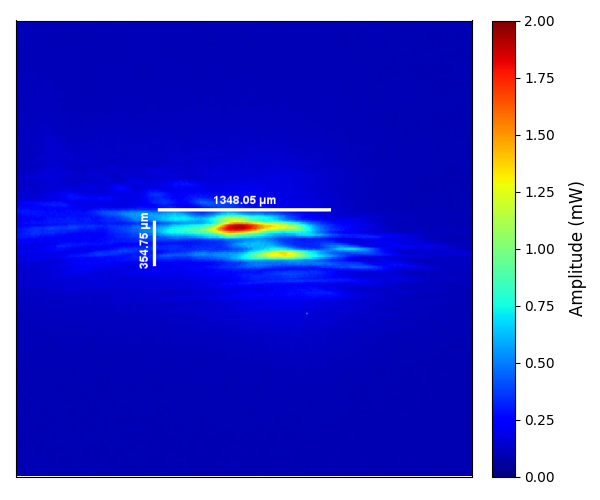}
        \caption{Spatial profiles obtained from diffraction on the amplitude spiral zone plate. 
        Left: optical vortices; 
        Right: converted mode obtained after the vortex beam passes through the cylindrical lens.}
        \label{fig:amp_szp_exp}
    \end{figure}

\begin{figure} [!ht]
        \centering
        \includegraphics[width=0.49\linewidth]{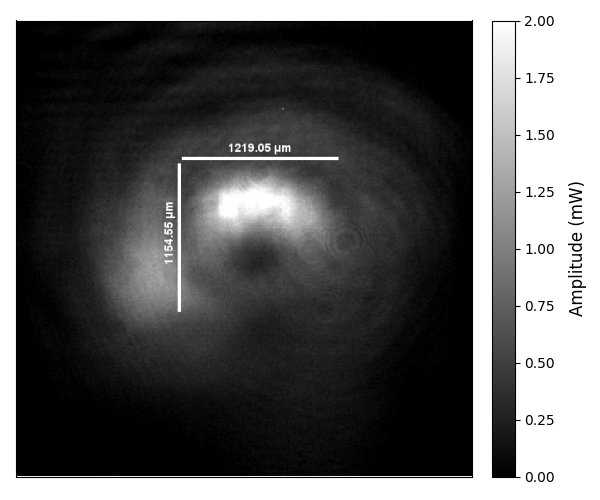}
        \includegraphics[width=0.49\linewidth]{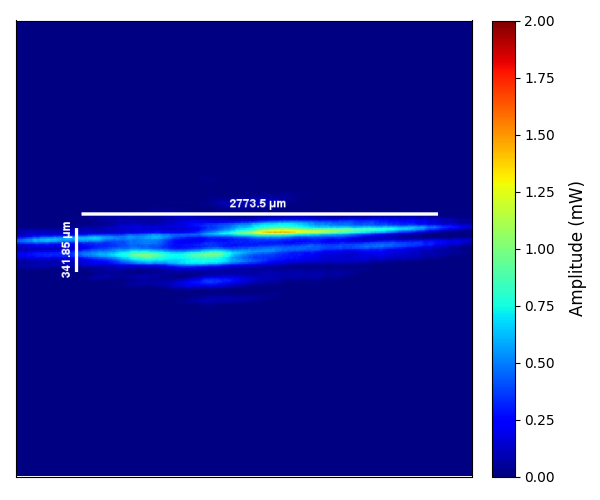}
        \caption{Spatial profiles obtained from diffraction on the phase spiral zone plate. 
        Left: optical vortices; 
        Right: converted mode obtained after the vortex beam passes through the cylindrical lens.}
        \label{fig:phase_szp_exp}
    \end{figure}

    Numerical simulations were first performed to validate the SZP designs, with results shown in Figs.~\ref{fig:sim_amp} and~\ref{fig:sim_phase}. 
    The experimental results are presented in Figures~\ref{fig:amp_szp_exp} and~\ref{fig:phase_szp_exp}. 
    For each figure, the left panel shows the raw intensity profile of the generated vortex beam, while the right panel shows the profile after passing through the cylindrical lens.

    In response to this reviewer’s comment, we have added amplitude scales to Figures~\ref{fig:sim_amp}, \ref{fig:sim_phase}, \ref{fig:amp_szp_exp}, and \ref{fig:phase_szp_exp}. 
    Additionally, we have included a detailed description of the experimental setup in Figure \ref{fig:exp_scheme}.



    For improved visualization of the converted mode structure, the images on the right were recorded using a false-color map. 
    This enhances the visual contrast of intensity gradients, making the separation between the stripes clearer than in a standard grayscale image. 
    The raw vortex profiles on the left were kept in grayscale to show the unmodified intensity distribution.

    The amplitude SZP results are shown in Figure~\ref{fig:amp_szp_exp}. 
    The generated vortex shows the expected annular intensity profile, though with noticeable noise and asymmetry. 
    The converted mode clearly displays two distinct lobes, corresponding to a topological charge of $\vert \ell \vert = 1$. 
    The quality is suboptimal, consistent with the theoretically lower efficiency of amplitude elements. The diffraction efficiency, defined as the ratio of power in the first-order vortex beam to the total incident power, was measured to be approximately 10\%.

    The phase SZP results are depicted in Figure~\ref{fig:phase_szp_exp}. 
    The vortex generated by the phase plate exhibits a more defined ring structure and less background noise. 
    Correspondingly, the converted mode shows two cleaner, more uniform fringes, again confirming a topological charge of $\vert \ell \vert = 1$. 
    The diffraction efficiency was approximately 40\%, a four-fold improvement over the amplitude plate, as expected from theory.

    Regarding discrepancies, the fringe orientation in the experimental results is opposite to that in the simulations, which is an artifact of the experimental setup, likely due to an inversion of the image by the optical train or camera sensor orientation, and does not affect the physical interpretation. 
    The vortex ring in Fig.~\ref{fig:amp_szp_exp} appears smaller than in Fig.~ \ref{fig:phase_szp_exp} because the fabricated phase SZP was designed with a much shorter focal length, resulting in a tighter focus, a difference that was intentional to explore fabrication constraints. 
    The experimental profiles exhibit some imperfections, which are attributed to the non-ideal Gaussian profile of the input laser beam and minor alignment inaccuracies, representing topics for future improvement.

\newpage
\section{Discussion and Conclusion}

    We have successfully designed, fabricated, and demonstrated amplitude and phase spiral zone plates (SZPs) for generating optical vortices directly in the deep ultraviolet (DUV) using a high-power photoinjector laser. 
    Vortex beams with a topological charge of $\ell = 1$ were generated, with phase SZPs achieving $\sim 40\%$ efficiency, significantly outperforming amplitude SZPs ($\sim 10\%$).

    \subsection{Advantages and Comparison with Other Methods}

    The key advantage of SZPs, especially for RF photoinjectors, lies in their ability to efficiently shape high-power DUV beams while withstanding high fluences—unlike SLMs. 
    Unlike up-conversion methods, our direct approach avoids nonlinear losses. 
    Though SPPs may offer higher theoretical efficiency, SZPs are easier to fabricate and scale, offering a better performance-to-complexity balance.

    \subsection{Error Analysis and Limitations}

    The vortex beam quality is affected by fabrication errors (e.g., etch depth deviations, edge roughness), input beam imperfections (e.g., astigmatism, spatial noise), and alignment sensitivity (e.g., SZP tilt). 
    These contribute to asymmetries and reduced mode purity. 
    Detailed mode purity analysis is a subject for future study.

    The reduced beam quality observed in the experimental profiles is primarily attributed to fabrication imperfections. 
    These arise both during photolithographic mask transfer—since the DUV operation wavelength requires zone structures with submicron spatial periods—and during plasma etching, where maintaining uniform depth over the entire area is challenging and requires localized process control. 
    Consequently, minor deviations in the groove depth and boundary definition lead to phase errors that degrade the purity of the generated vortex mode.

    \subsection{Comparison with Spiral Phase Plate}

    Table~\ref{tab:comparison} summarizes the main quantitative differences between the fabricated SZPs and conventional SPPs. 
    While SPPs can reach higher theoretical efficiencies (up to 80 -- 90\%), their continuous relief fabrication requires sub-nanometer surface precision and gray-scale lithography or diamond-turning processes, which are costly and time-consuming. 
    In contrast, binary phase SZPs achieve practical efficiencies of $\sim 40\%$ in the DUV and are fabricated via standard photolithography and plasma etching in a single step.

    \begin{table}[!htbp]
    \centering
    \caption{Comparison between SZPs and SPP} \label{tab:comparison} 
    \small
        \begin{tabular}{lccc}
        \hline
        \textbf{Parameter} & \textbf{SPP} & \textbf{Phase SZP} & \textbf{Amp. SZP} \\
        \hline
        Fabrication complexity & High & Moderate & Low \\
        DUV compatibility & Feasible, limited & Good & Good \\
        Efficiency & 80--90\% & $\sim$40\% & $\sim$10\% \\
        Cost & Very high & Low & Very low \\
        Damage threshold & High & High & Moderate \\
        \hline
    \end{tabular}
    \end{table}

\subsection{Future Outlook}

    This work lays the groundwork for DUV vortex generation. Future directions include refining fabrication for smaller features and extending designs to higher-order vortices ($l > 1$). 
    Studying their efficiency and quality is a priority.

    In summary, SZPs offer a practical, efficient route to structured light in the DUV. 
    Their integration into the JINR RF photoinjector opens the way for novel studies on relativistic electron vortex beams—benefiting accelerator physics, quantum state control, and advanced diagnostics.

\begin{figure} [!ht]
        \centering
        \includegraphics[width=0.4\linewidth]{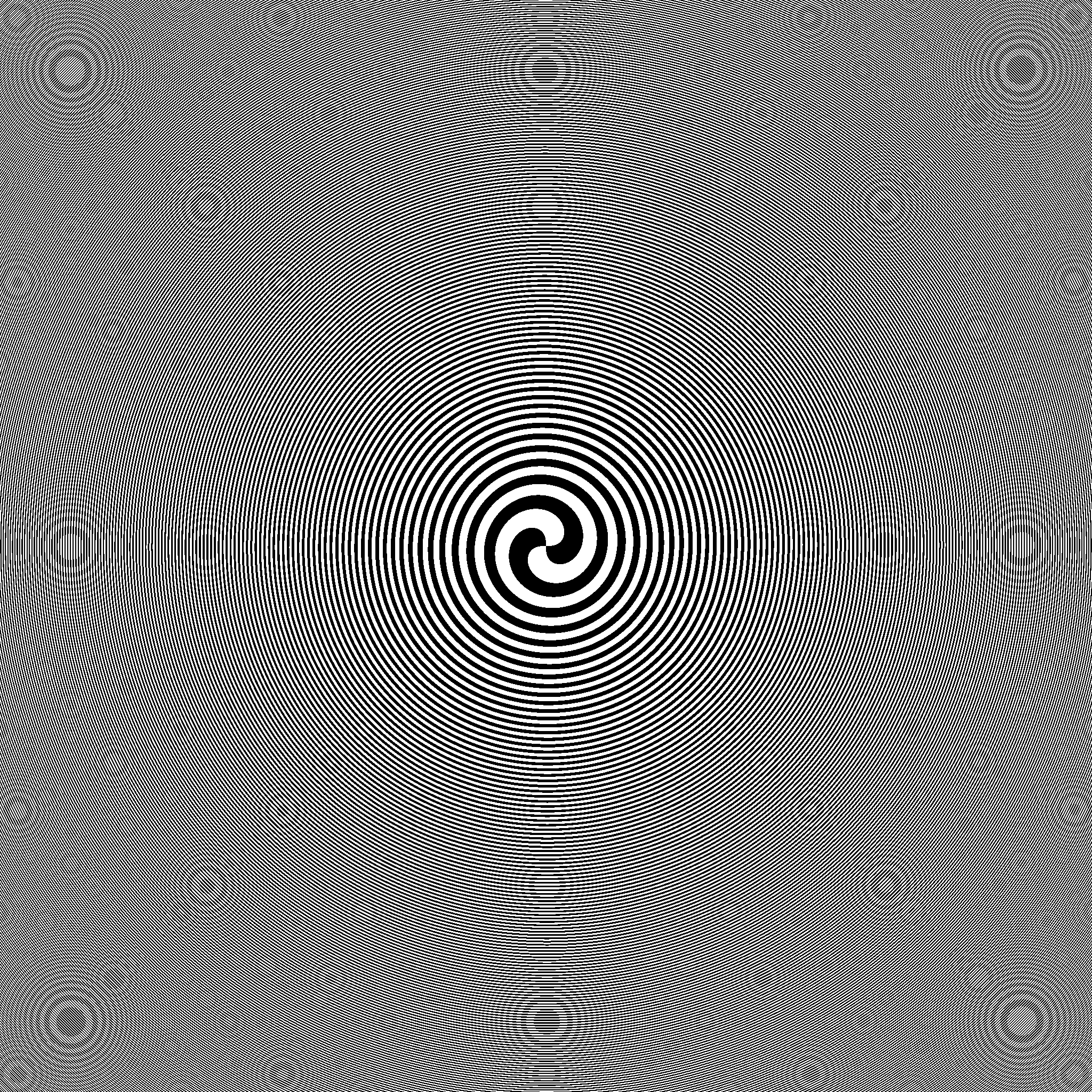}
        \includegraphics[width=0.4\linewidth]{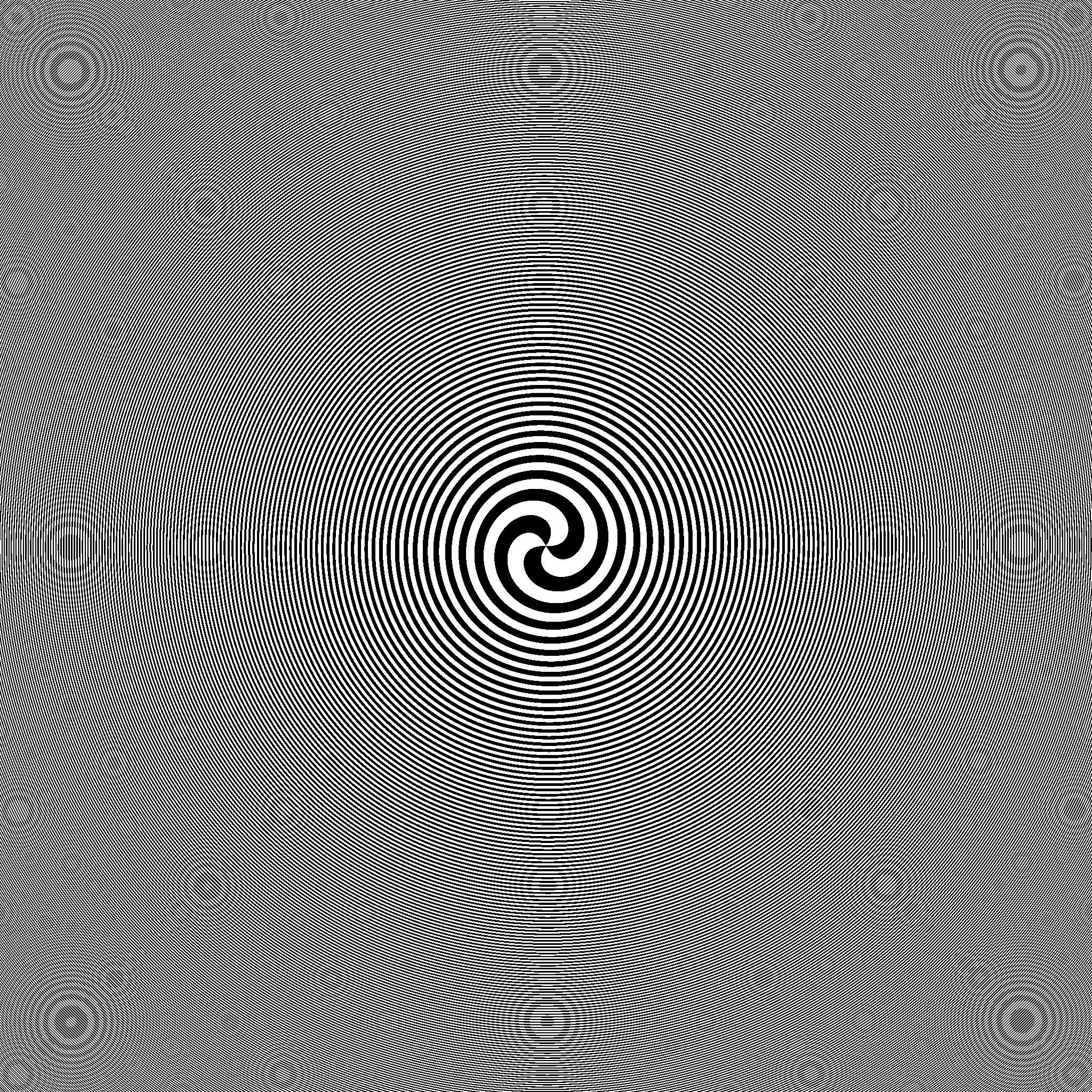}
        \caption{Generated SZPs masks for topological charges $\ell = 2$ and $\ell = 3$.}
        \label{fig:mask_l_2_3}
    \end{figure}

\begin{figure} [!ht]
        \centering
        \includegraphics[width=0.49\linewidth]{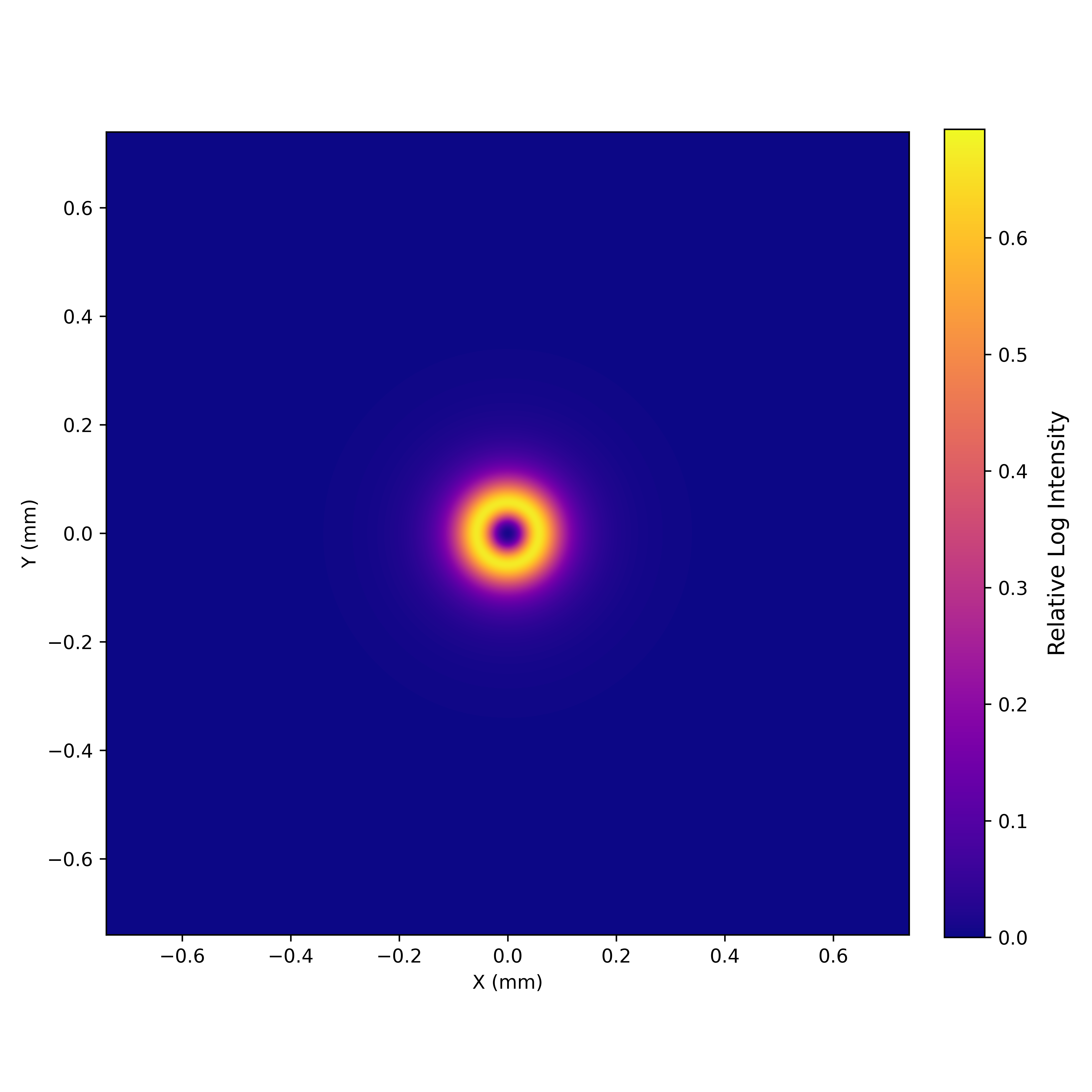}
        \includegraphics[width=0.49\linewidth]{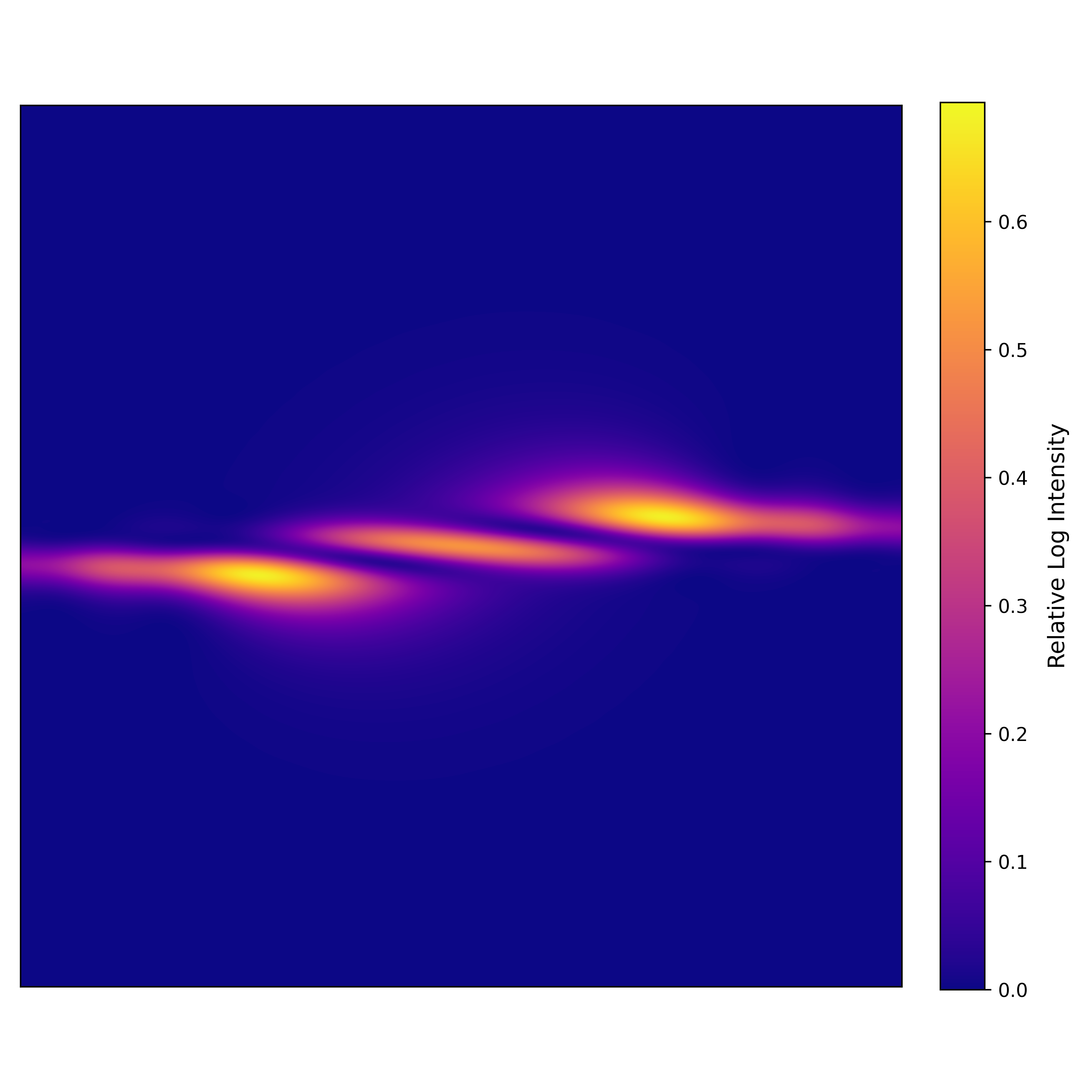}
        \caption{Simulated intensity distributions for diffraction on SZP with topological charges $\ell = 2$.}
        \label{fig:sim_l_2}
    \end{figure}

\begin{figure} [!ht]
        \centering
        \includegraphics[width=0.49\linewidth]{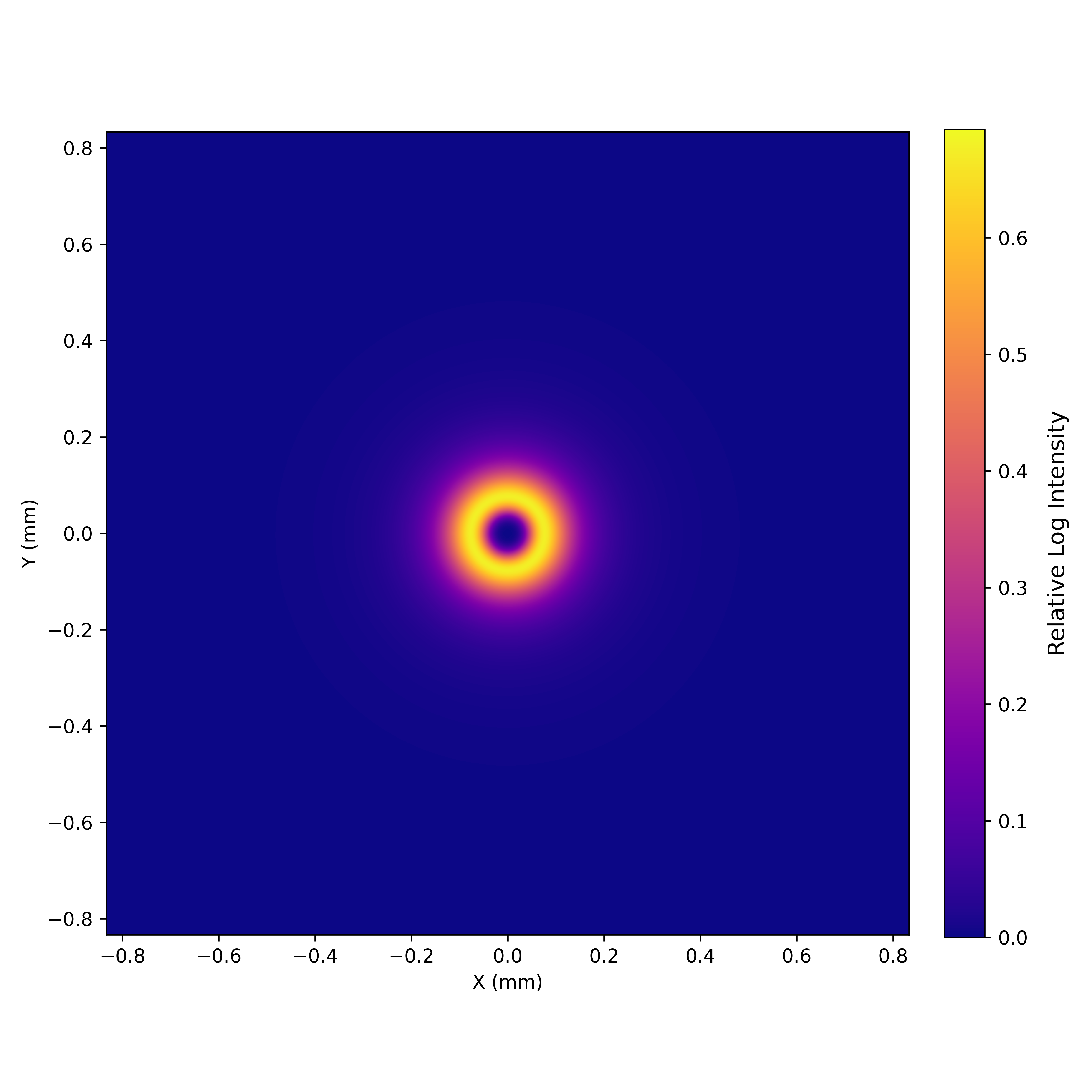}
        \includegraphics[width=0.49\linewidth]{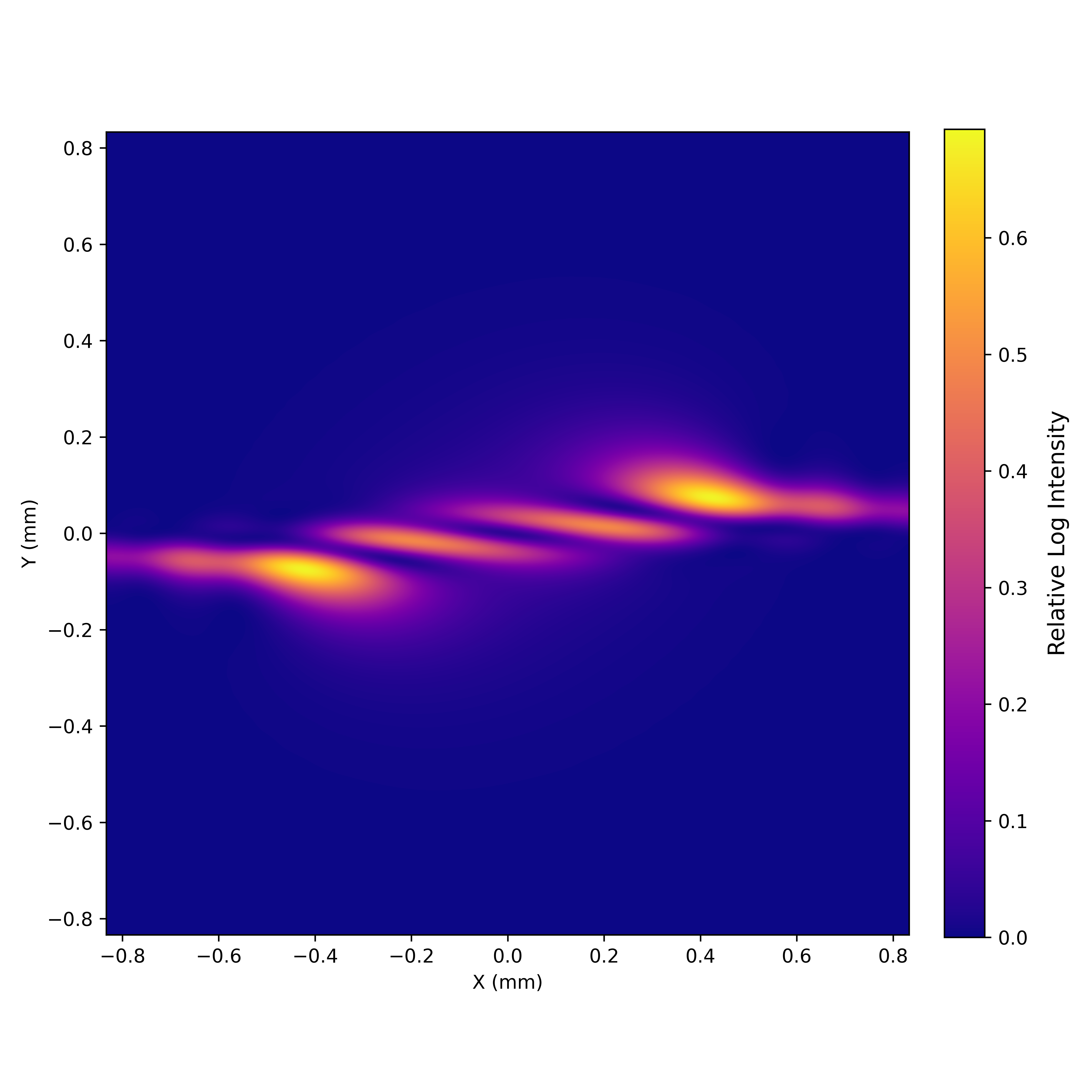}
        \caption{Simulated intensity distributions for diffraction on SZP with topological charges $\ell = 3$.}
        \label{fig:sim_l_3}
    \end{figure}

    Although this study focused on $l=1$ as a proof of concept, numerical simulations were also performed for higher-order topological charges ($l=2$ and $l=3$), confirming that the generated intensity distributions maintain the expected annular structures and mode conversion behavior (see Fig.~\ref{fig:mask_l_2_3},~\ref{fig:sim_l_2} and~\ref{fig:sim_l_3}). 
    However, due to the limited available laser power and beam size in the DUV range, experimental verification of higher-order SZPs is left for future work. 
    Fabrication of SZPs with $l>1$ does not introduce additional technological challenges, as it only modifies the spiral term in the design equation. 
    Importantly, the SZP fabrication technology easily scales to topological charges on the order of 8–10, beyond this range, challenges arise due to the decreasing feature size of the mask, which limits manufacturability.

\newpage
\section*{Aknowledgement}
    We sincerely thank K. Cherepanov for his invaluable assistance and advice. We are also grateful to S. Baturin for the helpful discussions and constructive feedback. 
    Special thanks to the Institute of Applied Physics of the Russian Academy of Sciences and M. Martyanov for rebuilding the drive laser and for their continued support in restoring it.    
    
    This study was supported by the Russian Science Foundation (Project No. 23-62-10026)~\cite{RSF}.

\bibliographystyle{elsarticle-num} 
\bibliography{refs}

\end{document}